\documentclass[preprint,amsmath,amssymb,prl]{revtex4}

\usepackage{graphicx}
\usepackage{dcolumn}
\usepackage{bm}

\begin{document}



\title{The quantum paraelectric behavior of $SrTiO_{3}$ revisited:\\
relevance of the structural phase transition temperature}

\author{Manuel I. Marqu\'es}
\email{manuel.marques@uam.es}
\author{Carmen Arag\'o}
\author{Julio A. Gonzalo}

\affiliation{%
~Departamento de F\'isica de Materiales C-IV,
Universidad Aut\'onoma de Madrid, 28049 Madrid, Spain\\
}%



\begin{abstract}
It has been known for a long time that the low temperature behavior shown by the dielectric
constant of quantum paraelectric $SrTiO_{3}$ can not be fitted properly by Barrett's formula
using a single zero point energy or saturation temperature ($T_{1}$). As it was originally shown 
[K. A. M\"{u}ller and H. Burkard, Phys. Rev. B {\bf 19}, 3593 (1979)] 
a crossover between two different
saturation temperatures ($T_{1l}$=77.8K and $T_{1h}$=80K) at $T\sim10K$ is needed to explain the
low and high temperature behavior of the dielectric constant. However, the physical reason for the 
crossover between these
two particular values of the saturation temperature at $T\sim10K$ is unknown.  
In this work we show that the
crossover between these two values of the saturation temperature at $T\sim10K$ can be taken as
a direct consequence of
(i) the quantum distribution of frequencies
$g(\Omega)\propto\Omega^{2}$ associated with the complete set of low-lying modes
and (ii) the existence of a definite maximum phonon frequency 
given by the structural transition critical temperature $T_{tr}$.
\end{abstract}


\maketitle


The "quantum paraelectrics" or "incipient ferroelectrics" have been a topic of considerable interest during
last decades. Quantum fluctuations associated to a non-negligible zero point energy prevent the onset of
ferroelectric long-range order. Actually, some non-proper ferroelectric perovskites such as Strontium Titanate
($SrTiO_{3}$) or Potassium Tantalate ($KTaO_{3}$), with very large dielectric contant values, undergo
ferroelectric phase transitions when doped with small amounts of polar impurities such as calcium or lithium
\cite{Bianchi}. Recently there has been some controversy about the origin of this quantum fluctuations. It is
not clear whether they come from a superposition of incoherent modes, or whether the coupling of the lowest
transverse acoustic and the soft mode gives rise to a coherent quantum paraelectric state \cite{Mullerintro}.
This conjecture has been stimulated by the observation of an anomaly, indicative of a possible phase transition
near $T_{c}=35K$ measured on several experimental studies based on light \cite{Vacher,Hehlen}, neutron scattering \cite{Courtens},
sound dispersion and attenuation \cite{Balashova, Nes}, EXAFS \cite{Fischer} and dielectric spectroscopy \cite{Viana,Hemberger}.

Almost thirty years ago M\"{u}ller and Burkhard \cite{Muller} measured the
dielectric constant ($\chi$) of $SrTiO_{3}$ in monodomain samples down to $T=0.3K$.
The experiment showed that $\chi$ was temperature independent below $T=3K$.
This constant region was in agreement with theoretical predictions from Barrett's formula \cite{Barrett}
which extended Slater's mean field theory of perovskites, to include quantum effects.
Barrett's formula depends on the saturation or Einstein temperature $T_{1}$, which is related to
the quantum mechanical zero point motion of the elementary dipoles ($T_{1}=\hbar\Omega/k$),
being $\Omega$ the single dipole frequency. However, M\"{u}ller et al. noted that Barrett's
formula was unable to fit $\chi(T)$ for all temperatures below $T=30K$ \cite{Muller}. Actually, two saturation
temperatures ($T_{1h}$ and $T_{1l}$) were needed to fit the $30K>T>10K$ region and
the $T<3K$ region
respectively \cite{Muller}. No single saturation temperature was able to fit the complete $30K>T>0.3K$ region.
This breaking from the experimental data was attributed to a coupling of the single soft mode to
other occupied acoustic modes \cite{Muller} considering phonon dressing \cite{Cowley} and dipolar
behavior \cite{Kind,Migoni} in order to reduce the classical critical temperature.
More recently, another approach to fit the experimental data for quantum paraelectrics in a large range
of temperatures has been proposed. It is based on a generalization of Barrett's formula to a quantum 
Curie-Weiss like formula
by introducing a critical exponent $\gamma$ \cite{Dec}. The obtained  value of $\gamma$ for
pure $SrTiO_{3}$ is $\gamma=1.7$.

Barrett's formula may be obtained rigorously \cite{Salje,SaljeII} by introducing a quantum temperature
scale related to the energy of the quantum oscillator

\begin {equation}
T_{q}=\frac{\hbar\Omega}{k}\left(\frac{1}{e^{\frac{\hbar\Omega}{kT}}-1}+\frac{1}{2}\right)
\end {equation}

This quantum temperature has been successfully applied to the analysis of the susceptibility in quantum paraelectrics
\cite{Kleemann} and, by means of a mean field approach, it has been also used to explain the quantum effects
found in mixed classical ferroelectric systems such as tris-sarcosine calcium chloride/bromide systems
\cite{Gonzalo,Arago}. Using this quantum temperature instead of the classical temperature in Slater's mean-field
formula for the susceptibility we obtain Barrett's formula for quantum paraelectrics:

\begin {equation}
\chi=\frac{C}{\frac{T_{1}}{2}\coth(\frac{T_{1}}{2T})-T_{c}}+A
\end {equation}

where $C$ is the Curie-Weiss constant, $T_{c}$ is the extrapolated Curie temperature of the ferroelectric transition and
$A$ is the temperature
independent part of the dielectric constant. Fitting parameters for $SrTiO_{3}$ are given by
$A=0$, $C=8\times10^{4}K$ and $T_{c}=35.5K$. As already mentioned, two saturation temperatures
$T_{1}=T_{1l}$ and $T_{1}=T_{1h}$ are needed to fit the low ($T<3K$) and high temperature regimes ($30K>T>10K$).
In particular, Muller et al. calculated that the best fists to the experimental data were found for $T_{1l}=77.8K$ and
$T_{1h}=80K$ \cite{Muller}. These two particular values are independent fitting parametres lacking a rigorous 
theoretical justification.

Very recently, Yuan et al. \cite{Wang} have modeled this crossover beteween $T_{1l}=77.8K$ and $T_{1h}=80K$  
by introducing 
a continuous, temperature dependent $T_{1}$, given by a hyperbolic tangent  
that fits very well dielectric experimental
results for $SrTiO_{3}$ in the whole range of temperatures. The proposed temperature dependence
is given by

\begin {equation}
T_{1}=T_{1l}+\frac{T_{1h}-T_{1l}}{2}\left[1+\tanh\left(\frac{T-\theta}{\alpha}\right)\right]
\end {equation}

where $\theta$ and $\alpha$ are two new free parameters introduced to describe the crossover temperature (close to $10K$)
and the crossover rate respectively. Note that four fitting parameters are needed to describe the crossover.

However, the reason for the existence of this crossover dependence on the saturation temperature is unknown. 
Why are the characteristic saturation temperatures associated with the quantum mechanical zero point energy of the 
elementary dipoles given by $T_{1l}=77.8K$ and $T_{1h}=80K$? Why is there a change on the dielectric behavior of the 
quantum paraelectric between these two particular values precisely at $T\sim10K$?

In the following, we will show how this crossover between this two particular saturation temperatures 
can be taken as due to the existence of the quantum distribution of frequencies
$g(\Omega)\propto\Omega^{2}$ associated
with the complete set of low-lying oscillations 
\cite{Rechester,Khmelnitskii} and to a specific maximum value for the phonon frequencyes given by the structural transition critical
temperature.

Let us consider the general case where, instead of $\Omega=cte$, the system presents a distribution
of frequencies given by $g(\Omega)$. Due to the existence of this distribution it is not appropriate to relate
the quantum temperature $T_{q}$ to
a single frequency $\Omega$ but to an average value ($<\Omega>$) which is temperature dependent.
The temperature dependence of the average frequency value is given by:

\begin {equation}
<\Omega>\left(\frac{1}{e^{\frac{\hbar<\Omega>}{kT}}-1}+\frac{1}{2}\right)=
\frac{\int_{0}^{\Omega_{max}}\Omega\left(\frac{1}{e^{\frac{\hbar\Omega}{kT}}-1}+\frac{1}{2}\right)g(\Omega)d\Omega}
{\int_{0}^{\Omega_{max}}g(\Omega)d\Omega}
\end {equation}

being $\Omega_{max}$ the maximum possible value for the frequency of the phonon oscillations at low
temperatures.

Once the frequency distribution of the system is fixed as $g(\Omega)\propto\Omega^{2}$, corresponding
to the low-lying modes, it is possible
to solve this equation numerically to obtain the temperature dependence of $<\Omega>$ and the behavior
of $T_{1}=\hbar<\Omega>/k$. In order to know the value of $\Omega_{max}$ we consider the following:
The structural cubic-tetragonal phase transition in $SrTiO3$ at $T_{tr}\sim104K$,
associated with an acoustic soft mode behavior \cite{Shirane}, should mark an upper boundary for the 
frequency of the oscillations
at low temperatures. Higher values of $\Omega$ should be unstable at these low temperatures.   
The maximum possible value for the frequency of the phonon oscillations is then given by $\Omega_{max}=T_{tr}k/\hbar$.

The value of $T_{tr}$ for $SrTiO_{3}$ depends somewhat on the sample\cite{Hunnefeld}, however most of the values found
in the literature are close to $T\sim104K$. So, for the
particular case studied in this work (sample A from \cite{Muller}) we take the
value $T_{tr}=104K$ (corresponding to $\Omega_{max}=13.6THz$).
Once this value for $\Omega_{max}$ is introduced in Eq.4, the existence of a crossover for $T_{1}(<\Omega>)$ 
between the two saturation temperatures ($T_{1l}=77.8K$ and $T_{1h}=80K$) appears as
a natural consequence of the existence of the frequency distribution $g(\Omega)$ of low-lying modes with 
$\Omega_{max}=T_{tr}k/\hbar$ 
(see Fig.1). Note how the crossover starting point at approximately $\theta=10K$ appears also straightforwardly.

\begin{figure}
\includegraphics[width=7cm,height=7cm,angle=0]{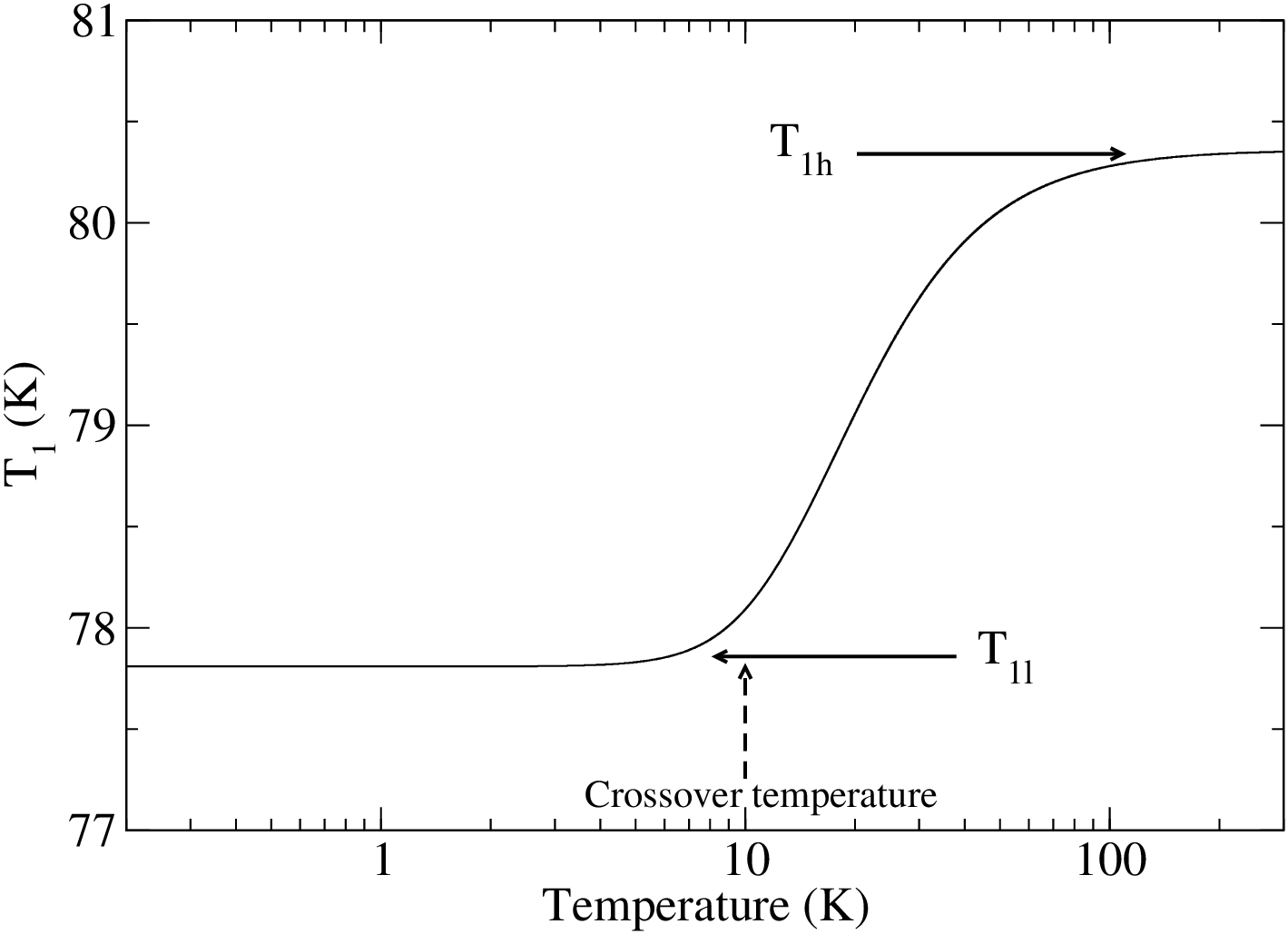}
\caption{Behavior of the saturation temperature of the system vs. temperature
}
\label{fig1}
\end{figure}

Of course, once we know the temperature dependence of $T_{1}$ it is possible to obtain
the behavior of the dielectric constant using Barrett's formula (Eq.2). The final result is shown in
Fig.2 together with the two fittings given by K.A.Muller et al. \cite{Muller} for the low and high temperature
regimes.
Note how the data between both regimes ($30K>T>10K$ and $T<3K$) is now explained by means of a single
distribution of frequencies in $SrTiO_{3}$ given by $g(\Omega)\propto\Omega^{2}$ with $\Omega_{max}=T_{tr}k/\hbar$.

This relation found between the structural critical temperature and the low temperature value of the dielectric constant,
allows determining the critical temperature with very high precision.
In order to get $T_{tr}$ we scan $\Omega_{max}$ 
to obtain successive
$T_{1l}$ values until we match the particular value $T_{1l}=77.8K$ given by Muller et al \cite{Muller} to fit the
dielectric constant in the low temperature region (sample A). In Fig.3 we present $\mid T_{1l}-77.8K\mid$ vs.
$\hbar\Omega_{max}/k$. Note how the difference is zero at $\hbar\Omega_{max}/k=103.7K$, which is very close
to the average critical value $T_{tr}\sim104K$ for SrTiO3. 

Note that, in this way, we have obtained the
structural transition critical temperature $T_{tr}=103.7K$
by means of an indirect method, using only data from low temperature dielectric constant measurements. Since the value
for $T_{1l}$ or, equivalently,
the dielectric constant at $T\sim0K$, is very sensitive to the value of  $\hbar\Omega_{max}/k$ this method
turns out to be a very precise way to determine the structural critical temperature of a particular sample (for
example, knowing that the value of the dielectric constant at zero temperature of sample B in \cite{Muller} 
is $20^{-3}$ we get a critical temperature $T_{tr}=105.3K$, slightly higher than the one
corresponding to sample A). 

To conclude, as it is well known, Barrett's formula is insufficient to explain the dielectric behavior in $SrTiO_{3}$ for $T<30K$ when
using a single, contant value, for the frequency associated with the quantum mechanical zero point motion of
the elementary dipoles. This problem is solved when considering a quantum distribution of frequencies
$g(\Omega)\propto\Omega^{2}$ associated with the complete set of low-lying oscillations which are dominant at
low temperatures and a upper frequency given by $\Omega_{max}=T_{tr}k/\hbar$. This distribution gives rise to 
the existance of a crossover between two saturation temperatures capable
of explaining satisfactorily the behavior of the dielectric constant at any temperature. Since, the maximum value 
for the frequency
in the $g(\Omega)$ distribution is given by the structural transition critical temperature, it is also possible
to obtain very precise values for the structural critical temperatures by using
the measured $T=0K$ dielectric constant of the quantum paraelectric.

Helpful comments from C.L. Wang are greatfully acknowledged. We acknowledge financial support 
through grant DGICyT, FIS2004-00437.

\begin{figure}
\includegraphics[width=7cm,height=7cm,angle=0]{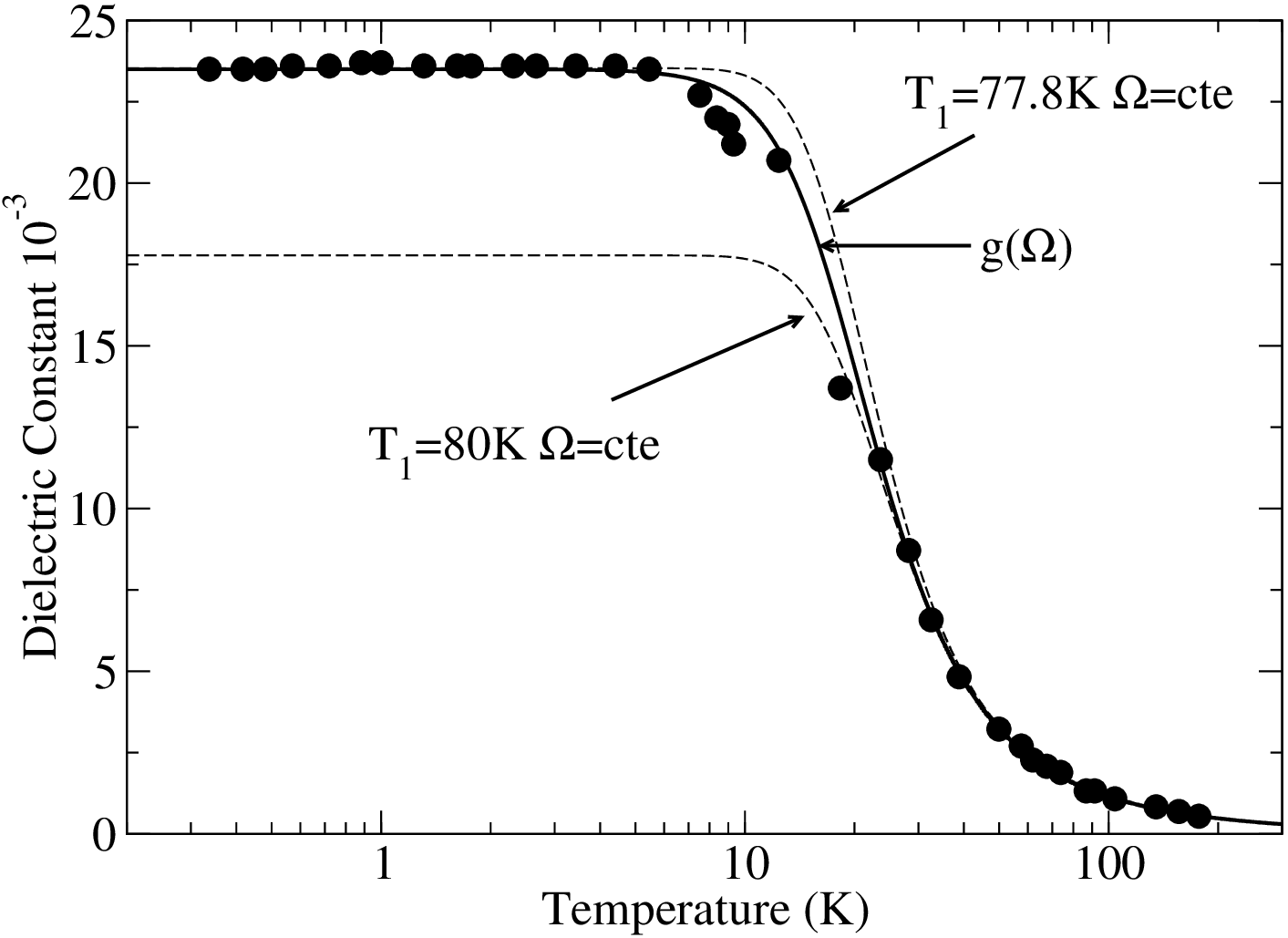}
\caption{Dielectric contant vs. temperature for $SrTiO_{3}$. Dotted lines are results from Barrett's formula
with $T_{1}=77.8K$ and $T_{1}=80K$ and using a constant value for the frequency associated with the
quantum mechanical zero point motion. Full line is the result obtained using a distribution
$g(\Omega)\propto\Omega^{2}$ and an structural critical temperature for the system equal to $104K$
}
\label{fig2}
\end{figure}

\begin{figure}
\includegraphics[width=7cm,height=7cm,angle=0]{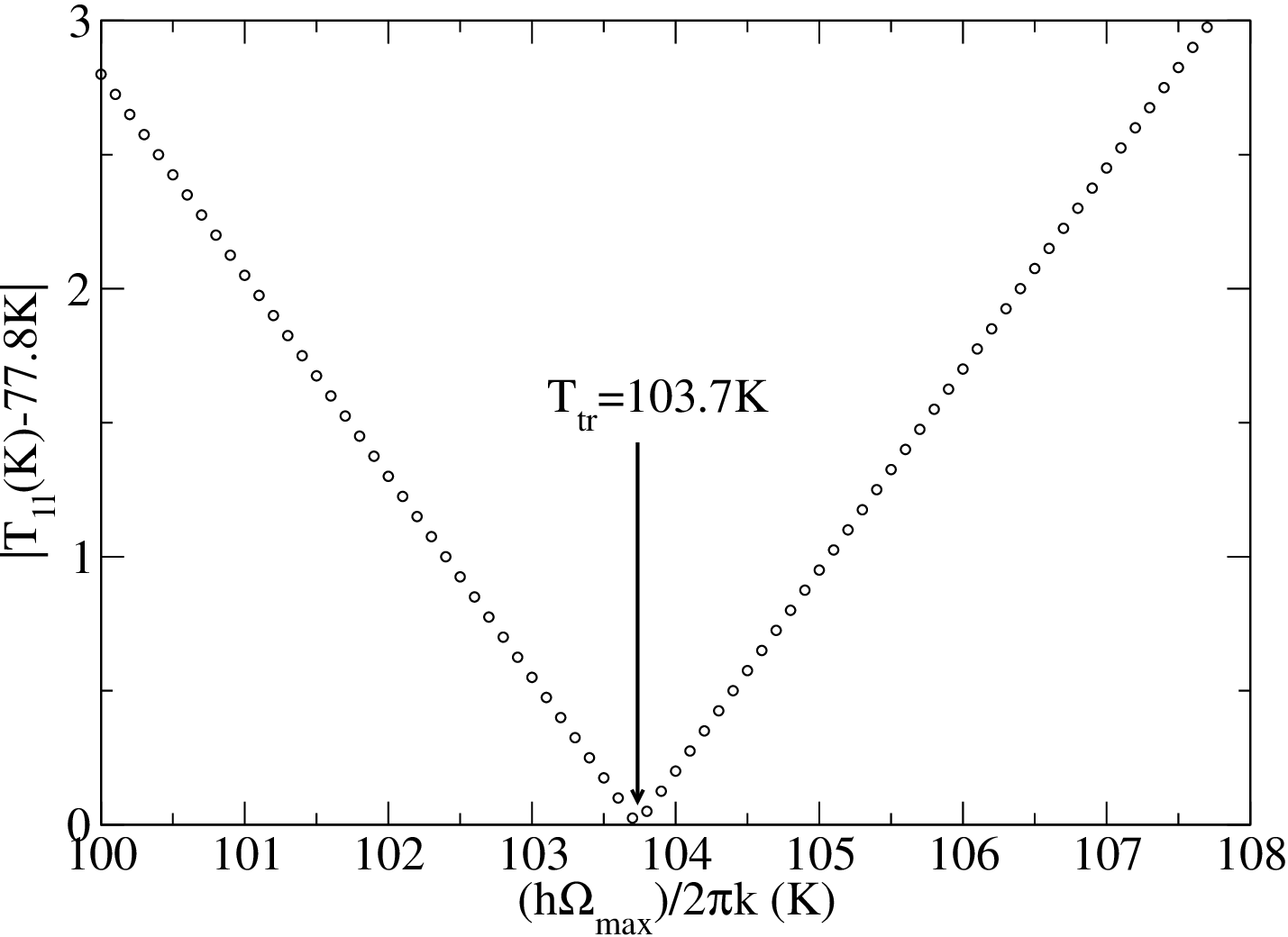}
\caption{Absolute value of the difference between the $T=0K$ saturation temperature ($T_{1l}$) and the saturation
temperature that fits the dielectric behavior at $T\sim0K$ \cite{Muller} vs. $\hbar\Omega_{max}/k$.
 We find that the value corresponding to this particular
$SrTiO_{3}$ system is $T_{tr}=103.7K$, which may be associated with the structural critical temperature of
the sample.
}
\label{fig3}
\end{figure}


\newpage

\begin{thebibliography}{99}

\bibitem{Bianchi}
See f.i. U. Bianchi, J. Dec, W. Kleeman and J. G. Bednorz, Phys. Rev. B {\bf 51}, 8737 (1995) and references therein.

\bibitem{Mullerintro}
K. A. M\"{u}ller, W. Berlinger and E. Tosatti, Z. Phys. B {\bf 48} 277, (1991).

\bibitem{Vacher}
R. Vacher, J. Pelous, B. Hennion, G. Coddens, E. Courtens and K. A. M\"{u}ller, Europhys. Lett. {\bf 17}, 45 (1992).

\bibitem{Hehlen}
B. Hehlen, A. -L. P\'{e}ron, E. Courtens and R. Vacher, Phys. Rev. Lett. {\bf 75} 2416, (1995).

\bibitem{Courtens}
E. Courtens, G. Coddens, B. Hennion, B. Hehlen, J. Pelous and R. Vacher, Phys. Scr. T {\bf 49A} 430 (1993).

\bibitem{Balashova}
E. V. Balashova, V. V. Lemanov, K. Kunze, G. Mart\'{i}n and M. Weihnacht, Solid State Commun. {\bf 94}, 17 (1995)

\bibitem{Nes}
O. M. Nes, K. A. M\"{u}ller, T. Suzuki and F. Fossheim, Europhys. Lett. {\bf 19}, 397 (1992).

\bibitem{Fischer}
M. Fischer, A. Lahmar, M. Maglione, A. San Miguel, J. P. Iti\'{e}, A. Polian and F. Baudelet, Phys. Rev. B {\bf 49}, 12451 (1994).

\bibitem{Viana}
R. Viana, P. Lunkenheimer, J. Hemberger, R. B\"{o}hmer and A. Loidl, Phys. Rev. B {\bf 50}, 601 (1994)

\bibitem{Hemberger}
J.Hemberger, M. Nicklas, R. Viana, P. Lunkenheimer, A. Loidl and R. B\"{o}hmer, J. Phys: Condens. Matter {\bf 8}, 4673 (1996).

\bibitem{Muller}
K. A. M\"{u}ller and H. Burkard, Phys. Rev. B {\bf 19}, 3593 (1979).

\bibitem{Barrett}
J. H. Barrett, Phys. Rev. {\bf 86}, 118 (1952).

\bibitem{Cowley}
R. A. Cowley, Philos. Mag. {\bf 11}, 673 (1965).

\bibitem{Kind}
R. Kind and K. A. M\"{u}ller, Comm. Phys. {\bf 1}, 223 (1976).

\bibitem{Migoni}
R. Migoni, H. Bilz and D. B\"{a}uerle, Phys. Rev. Lett. {\bf 37}, 1155 (1976).

\bibitem {Dec}
J. Dec and W. Kleemann, Solid State Commun. {\bf 106}, 695 (1998).

\bibitem {Kleemann}
W. Kleemann, J. Dec and B. Westwanski, Phys. Rev. B {\bf 58}, 8985 (1998).

\bibitem{Salje}
E. K. H. Salje, B. Wruck and H. Thomas, Z. Phys. B-Condens. Matter, {\bf 82}, 399 (1991).

\bibitem{SaljeII}
E. K. H. Salje, B. Wruck and S. Marais, Ferroelectrics, {\bf 124}, 185 (1991).

\bibitem{Gonzalo}
J. A. Gonzalo, Ferroelectrics, {\bf 168}, 1 (1995).

\bibitem{Arago}
C. Arag\'{o}, J. Garc\'{i}a, J. A. Gonzalo, C. Wang, W. Zhong and X. Xue, Ferroelectrics, {\bf 301}, 113 (2004).


\bibitem{Wang}
M. Yuan, C. L. Wang, Y. X. Wang, R. Ali and J. L. Zhang, Solid State Comm. {\bf 127}, 419 (2003). 

\bibitem{Rechester}
A.B. Rechester, Zh. \'{E}ksp. Teor. Fiz. {\bf 60} (2), 782 (1971)
[Sov. Phys. JETP {\bf 33}, 423 (1971)].

\bibitem{Khmelnitskii}
D. E. Khmel'nitskii and V. L. Shneerson, Fiz. Tverd. Tela (Leningrad) {\bf 13} (3), 832 (1971)
[Sov. Phys.- Solid State {\bf 13}, 687 (1971)].

\bibitem{Shirane}
G. Shirane, Rev. Mod. Phys. {\bf 46}, 437 (1974).

\bibitem{Hunnefeld}
H. H\"{u}nnefeld, T. Niem\"{o}ller, J. R. Schneider, U. R\"{u}tt, S. Rodewald, J. Fleiq and G. Shirane, Phys. Rev. B, {\bf 66}, 014113 (2002).



\end{thebibliography}
\end{document}